\documentclass[aps,prl,twocolumn,showpacs,superscriptaddress]{revtex4-1}
\usepackage{amssymb}
\usepackage{amsmath}
\usepackage{textcomp}
\usepackage{graphicx} 
\usepackage{dcolumn}
\usepackage{bm}
\usepackage{upgreek}
\usepackage{color}

\def\um{\mu\mbox{m}}

\def\Wcm2{\mbox{W cm}^{-2}}
\def\Wcmum2{\mbox{Wcm}^{-2}\mu\mbox{m}^{2}}

\def\cm3{\mbox{cm}^{-3}}

\begin{document}

\title{Beamed neutron emission driven by laser accelerated light ions}

\author{S.~Kar}\email{s.kar@qub.ac.uk}
\affiliation{Centre for Plasma Physics, School of Mathematics and Physics, Queen's University Belfast, BT7 1NN, UK}

\author{A.~Green}
\affiliation{Centre for Plasma Physics, School of Mathematics and Physics, Queen's University Belfast, BT7 1NN, UK}

\author{H.~Ahmed}
\affiliation{Centre for Plasma Physics, School of Mathematics and Physics, Queen's University Belfast, BT7 1NN, UK}

\author{A.~Alejo}
\affiliation{Centre for Plasma Physics, School of Mathematics and Physics, Queen's University Belfast, BT7 1NN, UK}

\author{A.P.L.~Robinson}
\affiliation{Central Laser Facility, Rutherford Appleton Laboratory, Didcot, Oxfordshire, OX11 0QX, UK}

\author{M.~Cerchez}
\affiliation{Institut f\"ur Laser-und Plasmaphysik, Heinrich-Heine-Universit\"at, D\"usseldorf, Germany}

\author{R.~Clarke}
\affiliation{Central Laser Facility, Rutherford Appleton Laboratory, Didcot, Oxfordshire, OX11 0QX, UK}

\author{D.~Doria}
\affiliation{Centre for Plasma Physics, School of Mathematics and Physics, Queen's University Belfast, BT7 1NN, UK}

\author{S. Dorkings}
\affiliation{Central Laser Facility, Rutherford Appleton Laboratory, Didcot, Oxfordshire, OX11 0QX, UK}

\author{J. Fernandez}
\affiliation{Central Laser Facility, Rutherford Appleton Laboratory, Didcot, Oxfordshire, OX11 0QX, UK}
\affiliation{Instituto de Fusi\'on Nuclear, Universidad Polit\'ecnica de Madrid, 28006 Madrid, Spain}

\author{S.R. Mirfyazi}
\affiliation{Centre for Plasma Physics, School of Mathematics and Physics, Queen's University Belfast, BT7 1NN, UK}

\author{P.~McKenna}
\affiliation{Department of Physics, SUPA, University of Strathclyde, Glasgow G4 0NG}

\author{K. Naughton}
\affiliation{Centre for Plasma Physics, School of Mathematics and Physics, Queen's University Belfast, BT7 1NN, UK}

\author{D.~Neely}
\affiliation{Central Laser Facility, Rutherford Appleton Laboratory, Didcot, Oxfordshire, OX11 0QX, UK}

\author{P.~Norreys}
\affiliation{Central Laser Facility, Rutherford Appleton Laboratory, Didcot, Oxfordshire, OX11 0QX, UK}
\affiliation{Rudolf Peierls Centre for Theoretical Physics, University of Oxford, Oxford, OX1 3NP, UK}

\author{C. Peth}
\affiliation{Institut f\"ur Laser-und Plasmaphysik, Heinrich-Heine-Universit\"at, D\"usseldorf, Germany}

\author{H.~Powell}
\affiliation{Department of Physics, SUPA, University of Strathclyde, Glasgow G4 0NG}

\author{J.A. Ruiz}
\affiliation{Colegio Los Naranjos, Fuenlabrada, 28941, Madrid, Spain}

\author{J. Swain}
\affiliation{Rudolf Peierls Centre for Theoretical Physics, University of Oxford, Oxford, OX1 3NP, UK}

\author{O.~Willi}
\affiliation{Institut f\"ur Laser-und Plasmaphysik, Heinrich-Heine-Universit\"at, D\"usseldorf, Germany}

\author{M.~Borghesi}
\affiliation{Centre for Plasma Physics, School of Mathematics and Physics, Queen's University Belfast, BT7 1NN, UK}

\date{\today}

\begin{abstract}

We report on the experimental observation of beam-like neutron emission with peak flux of the order of 10\textsuperscript{9} n/sr, from light nuclei reactions in a pitcher-catcher scenario, by employing MeV ions driven by high power laser. The spatial profile of the neutron beam, fully captured for the first time by employing a CR39 nuclear track detector, shows a FWHM divergence angle of $\sim70^{\circ}$, with a peak flux nearly an order of magnitude higher than the isotropic component elsewhere. The observed beamed flux of neutrons is highly favourable for a wide range of applications, and indeed for further transport and moderation to thermal energies. A systematic study employing various combinations of pitcher-catcher materials indicates the dominant reactions being d(p, n+p)\textsuperscript{1}H and d(d,n)\textsuperscript{3}He. 
Albeit insufficient cross-section data are available for modelling, the observed anisotropy in the neutrons' spatial and spectral profiles are most likely related to the directionality and high energy of the projectile ions.

\end{abstract}

\pacs {}

\maketitle


Neutrons provide many opportunities for probing of materials in ways which charged particles and ionizing radiation cannot. In this context an ultrashort, directional burst of neutrons with high flux in the MeV range would have wide ranging applications. Exciting opportunities for ultrafast studies lie in the area of materials for fusion energy research due to growing interest in understanding neutron-induced damage at the atomic scale~\cite{Perkins}. 
Furthermore, a compact source of pulsed, MeV neutrons would provide novel capabilities for interrogation of large cargo containers by fast neutron radiography techniques~\cite{NRS,PFNA1}, where the nature and location of the threat can be identified by simultaneously measuring scattered neutrons and time of flight of the induced gamma radiation.

Significant attention has been paid recently to laser driven sources capable of producing short neutron bursts, and having potential advantages in terms of cost reduction and compactness, reduction of radioactive pollution and ability of radiation confinement by close-coupled experiments. Although a different approach (high energy deuteron-breakup) has recently been reported\cite{Roth_Neutron}, the most established route to create a laser based neutron source is by employing laser accelerated ions in either fusion or spallation reactions. Since spallation of heavy atoms requires high energy projectile ions, reactions based on low atomic mass nuclei, such as protons, deuterons, lithium etc., 
are particularly relevant. 
The neutron yield from nuclear reactions scales with the product of the densities of the interacting species and the cross-section $\sigma$, which for most common reactions reaches high values at $\sim$MeV centre-of-mass energy. Producing high fluxes of MeV ions using intense lasers is currently a very active area of research. Where a number of emerging ion acceleration mechanisms, such as radiation pressure acceleration (RPA) \cite{RPA} and breakout afterburner (BOA) \cite{BOA}, hold the promise for producing higher energy ions with higher efficiency, target normal sheath acceleration (TNSA)~\cite{TNSA} is a well-established and robust mechanism which produces MeV ions with high flux and narrow divergence. Such beams can be readily deployed in a pitcher-catcher setting for neutron generation via nuclear reactions.

\begin{figure}
\includegraphics[angle=-90,width=0.48\textwidth]{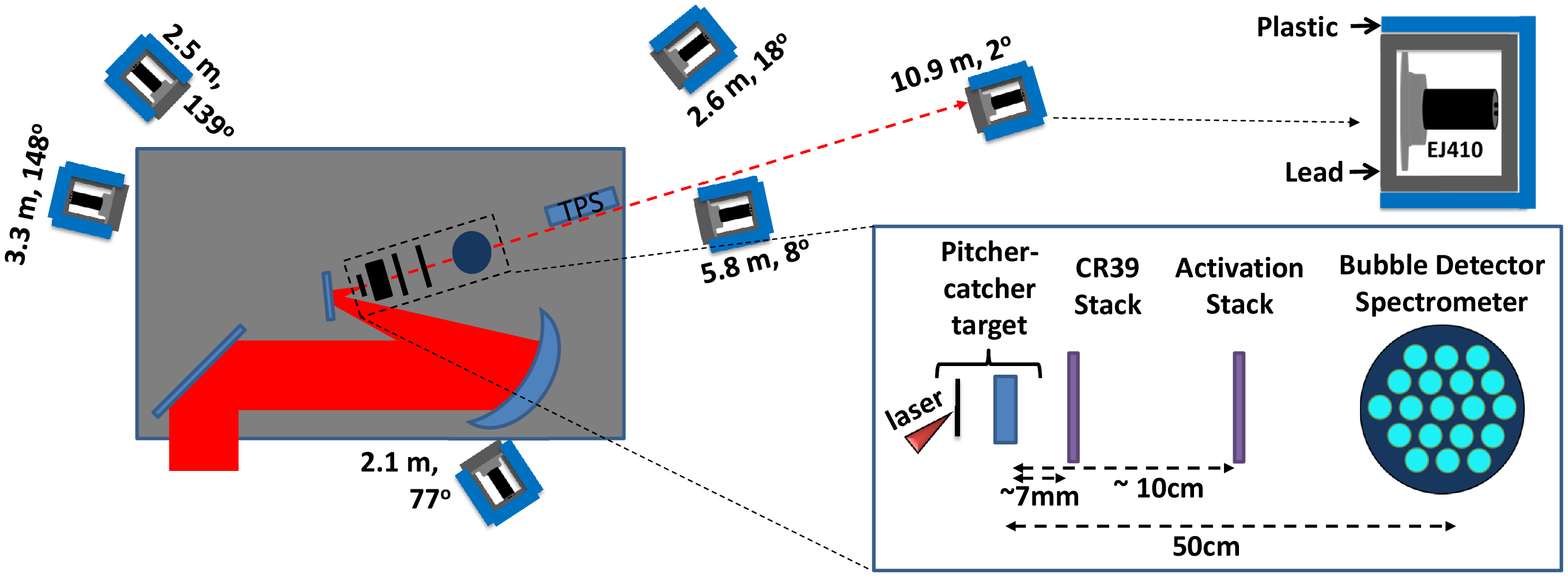}
\caption{The experimental set-up showing the in-chamber diagnostics and arrangements of nToF detectors around the chamber.}
\label{Setup}
\end{figure}

\begin{figure*}
\includegraphics[angle=-90,width=0.95\textwidth]{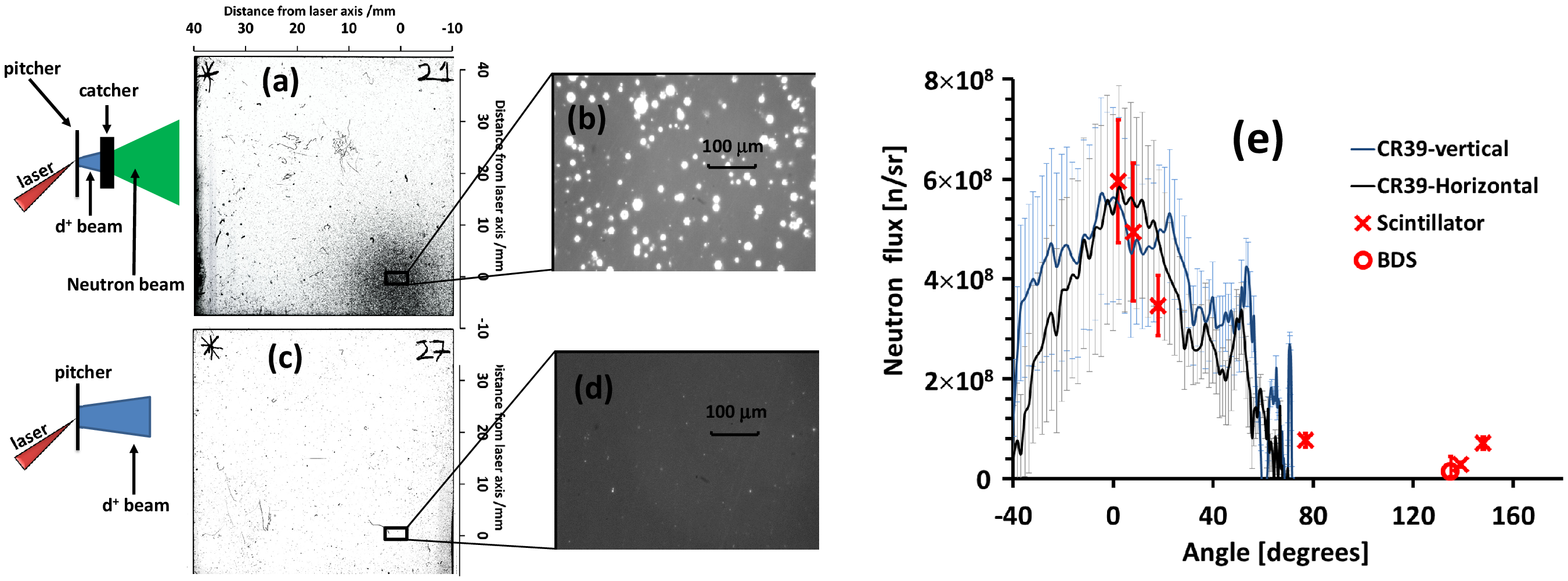}
\caption{The neutron beam can be clearly seen in the lower right of the CR39 slide in \textbf{(a)}. A close-up of the pits created on the CR39 can be seen in \textbf{(b)}, obtined by using a transmission microscope with 10x zoom, which can be compared with the \textbf{(d)}, in which there is no catcher target to convert the ions into neutrons. The laser axis was 1 cm up from the bottom of the CR39, and 1 cm in from the right hand edge. \textbf{(e)} shows the horizontal and vertical beam profiles from the scanned CR39 slide, along with the neutron flux (between 2.5-10 MeV) obtained from the nToF and bubble detector data from the same shot.}
\label{cr39}
\end{figure*}

In addition to the benefit of high reaction cross-sections, anisotropy in neutron emission is another facet of beam-nuclear reactions. 
Simulations~\cite{Petrov_sim} show that using several MeV ions in the above mentioned reaction involving low Z materials yields a neutron flux strongly peaked along the ion beam forward direction, and that the anisotropy grows further with increase in ion beam energy. A beamed neutron source is highly favourable not only for the aforementioned range of applications, but also for its further transport and an efficient moderation to thermal energies required for another range of applications~\cite{isis}, including for instance, Boron Neutron Capture Therapy~\cite{bnct}. Whilst some degree of neutron beam anisotropy has been recently reported at low neutron fluxes~\cite{Williangle_2011, Zulick_2013}, here we show, for the first time, direct imaging of a true beam-like neutron emission with peak flux of the order of $10^9$ n/sr. In particular, a neutron beam with a FWHM of $(70\pm10)^{\circ}$ and peak flux of (5 $\pm$ 2) x 10\textsuperscript{8} n/sr was captured in nuclear track detector kept in the proximity to the source. The neutron beam was produced by d(p, n+p)\textsuperscript{1}H and d(d,n)\textsuperscript{3}He reactions driven in a beam-catcher scenario, by employing protons and deuterons from thin deuterated plastic (CD) foils irradiated by a sub-petawatt laser. 



The experiment was performed using the petawatt arm of the Vulcan laser at the Central Laser Facility of STFC, UK. The linearly polarised laser pulse of 1053 nm wavelength, delivering $\sim$200 J on target after reflection from a plasma mirror, was focused down to a focal spot of $\sim$6~$\mu$m FWHM, providing peak intensity on the target $\sim3\times10\textsuperscript{20}$ Wcm$\textsuperscript{-2}$. Various targets were irradiated by the laser in order to generate energetic ions via the TNSA mechanism, namely gold foils (10 $\um$ thick), 98\% deuterated polyethylene $(C_2 D_4)_n$ (henceforth called CD) foils with and without few microns thick Al foil at the rear side. The ion beams were diagnosed in the earlier part of the experiment using a high resolution Thomson parabola spectrometer (TPS)~\cite{Deborah_TP} with image plate detectors looking at the target normal direction, as shown schematically in Fig.~\ref{Setup}. 
Since traces for ions with the same charge to mass ratio overlap in the TPS, we implemented the differential filtering technique described by Alejo \emph{et al.}~\cite{Aaron_TP} in order to extract the deuteron spectra from the diagnostic. 
$\sim$2 mm thick solid blocks of CD and graphite were placed $\sim 3$ mm behind the pitcher target 
(henceforth called catcher) in order to generate neutrons from nuclear reaction. The transverse size of the catcher was large enough to capture the entire ion beam. A full suite of neutron diagnostics was deployed in order to diagnose the spatial and spectral profiles of the neutrons generated in different shots. In order to capture the flux profile of the emitted neutrons over a large solid angle, and due to its low detection efficiency, of the order of $10^{-4}$, the CR39 nuclear track detector was placed in close proximity (7 mm) to the catcher. The CR39 was shielded by 4.5 mm thick lead in order to stop the high energy protons (upto 50 MeV~\cite{srim}) produced at the pitcher target from reaching the CR39. The absolute neutron flux was obtained from the CR39 by using the etching method and calibration given by Frenje \emph{et al.}\cite{Frenje}. Absolutely calibrated Bubble Detector Spectrometers (BDS)~\cite{Bubbles} and nuclear activation diagnostics~\cite{activation} were used, behind the CR39 along the ion beam forward direction, at a distance of 10 cm and 50 cm respectively from the pitcher target. Since the BDS 
provides absolute neutron flux in six discrete energy intervals within the energy range from 0.01 MeV to 20 MeV, 
it is possible to ascertain the flux of MeV neutrons generated in the catcher by discounting the large signal produced by the lower energy, scattered neutrons bouncing within the target area and hitting the detector several times over a period of time. 
Using 1 mm thick pure indium foils and measuring the decay of $\textsuperscript{115}$In and $\textsuperscript{116}$In, the neutron flux was estimated within two energy intervals, namely 0-3 MeV and 0.7-15 MeV. 
Six neutron time-of-flight (nToF) detectors consisting of EJ410 plastic scintillators, optically-coupled to XP3330 photomultiplier tubes (PMT), were used to provide spectral information at different emission angles by the time-of-flight method. The detectors were shielded appropriately (typically by 5-10 cm of lead on every side and 5 cm of plastic all around except the front of the detector) in order to suppress saturation of signal from Bremsstrahlung radiation and to reduce the noise in the signal due to scattered low energy neutrons reaching the detector. The angle of observation and distance of each nToF detector is given in Fig.~\ref{Setup}. The nToF detectors were cross-calibrated~\cite{ntof_calibration} against the spectra obtained from the BDS, which were absolutely calibrated by the company\cite{Bubbles}.

\begin{figure}
\includegraphics[angle=-90,width=0.48\textwidth]{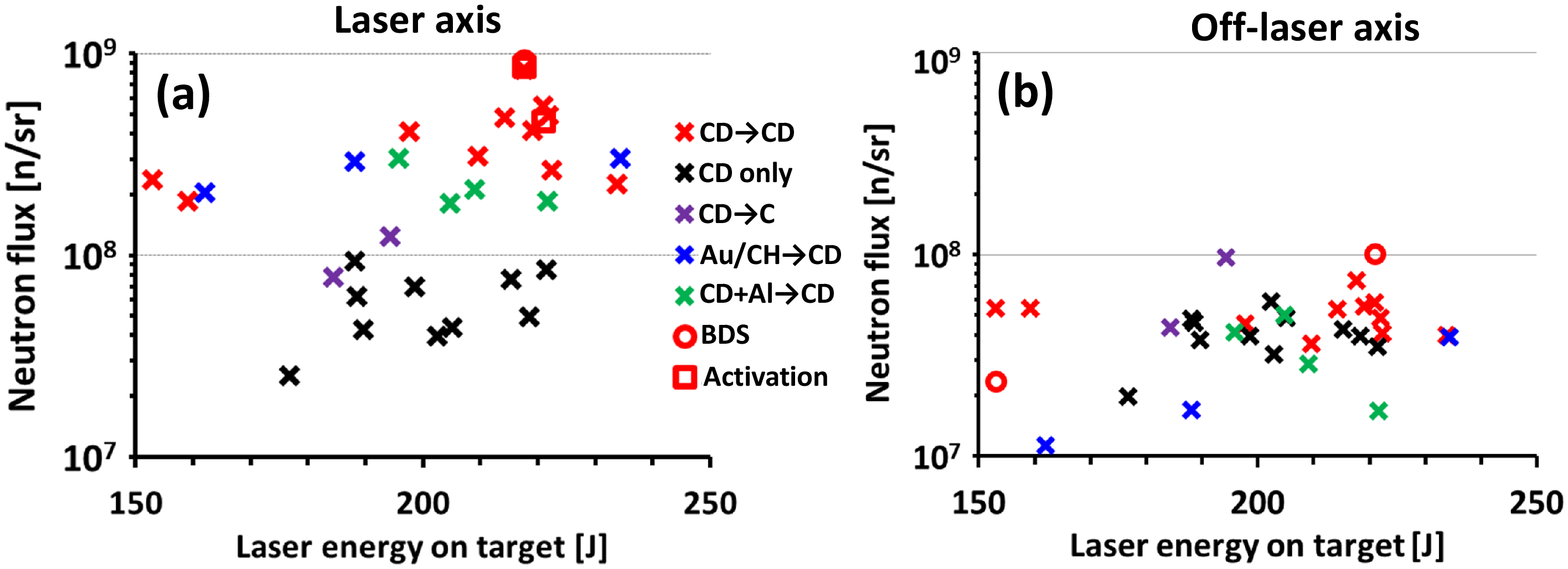}
\caption{\textbf{(a)} and \textbf{(b)} show the neutron flux (from 2.5 MeV to 10 MeV) along (a) the laser axis (average of the nToF detectors at 2$^\circ$ and 8$^\circ$) and (b) off laser axis (average of the nToF detectors at 77$^\circ$, 139$^\circ$ and 148$^\circ$) obtained for different pitcher and catcher materials and diagnostics. Where the color of the data points represent the pitcher and catcher material combinations, as mentioned in the figure legend, '$\times$', square and circle represent the data obtained by scintillators, activation and BDS respectively.}
\label{scan}
\end{figure}


As it can be clearly seen in Fig.~\ref{cr39}(a), the CR39 shows a beamed neutron emission from the catcher along the ion beam forward direction. The (0,0) co-ordinate in this image represents the ion beam axis, which is near the bottom right corner of the CR39. 
For comparison one can see Fig.~\ref{cr39}(c) showing the CR39 from the reference shot taken without the catcher, while keeping everything else the same. Both shots had identical pitcher targets of 10 $\mu$m CD and measured laser energy to within 1\% of each other. As expected, the amount of neutron generated in the later case (due to the interaction of ions, electrons and gamma rays produced from the pitcher target with the surrounding objects, including the lead shielding of the CR39) was not significant and lower than the detection threshold of the CR39. 
Neutrons are diagnosed in the CR39 due to the latent tracks created by the knock-on protons, which are revealed after etching in an alkali solution. The etched pits produced in the CR39 can be seen in the zoomed view of the CR39s, shown adjacent to the respective images. Using the calibration for detection efficiency of CR39 given by Frenje et. al.~\cite{Frenje}, the pit density across the CR39 was converted into neutron flux, as plotted in Fig.~\ref{cr39}(e). The horizontal and vertical lineouts of neutron flux across the CR39, passing through the co-ordinate (0,0), show an axisymmetric neutron flux profile with a FWHM divergence of $(70\pm10)^{\circ}$. The peak neutron flux along the ion beam axis was estimated as $(5\pm2)\times10\textsuperscript{8}$ n/sr. 

\begin{figure}
\includegraphics[angle=-90,width=0.48\textwidth]{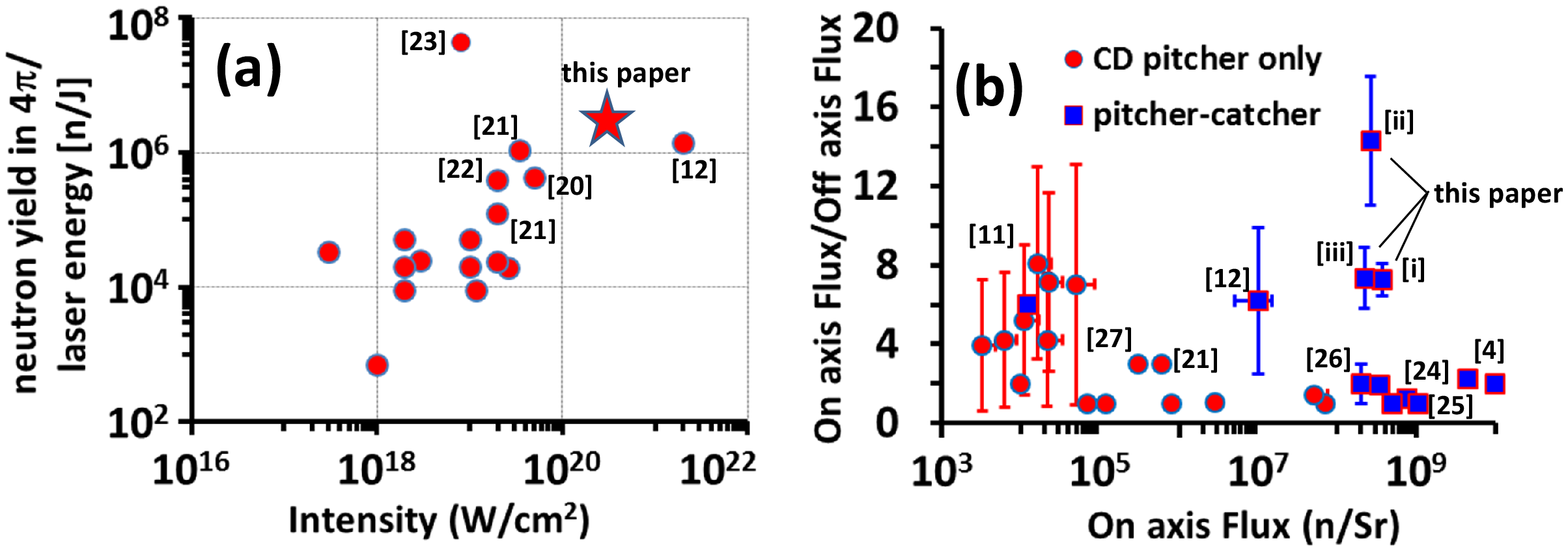}
\caption{(a) Graph showing neutron yield from pitcher-only configuration (red circles) using CD targets reported in literature, as cited beside the data points. The red star represents the average neutron yield from 10 $\um$ CD target obtained in our experiment, as shown in Fig.~\ref{scan}. (b) Graph showing the ratio between on and off axis neutron fluxes reported in literature, as cited beside the data points, with respect to on axis flux obtained in experiments. The data point obtained from our experiment is labelled as [i], [ii] and [iii] which corresponds to different pitcher and catcher combinations, such as CD$\rightarrow$CD, Au/CH$\rightarrow$CD and CD+Al$\rightarrow$CD respectively. The Au/CH$\rightarrow$CD case produced the highest degree of beam anisotropy, which is expected due to the low level of isotropic neutron flux produced from the non-deuterated pitcher targets, as can be seen from Fig.~\ref{scan}.}
\label{scaling}
\end{figure}

In order to identify the dominant nuclear reactions producing the beamed flux of forwardly directed neutrons, a systematic study was carried out by varying different pitcher and catcher materials. As can be seen in Fig.~\ref{scan}, the pitcher-only configuration using 10$\um$ CD target generates a fairly isotropic neutron emission in 4$\pi$ with an average flux $\sim 5\times 10^7$ n/sr. This corresponds to a total neutron yield in excess of $10^8$ neutrons ($>10^6$ neutrons/Joule), which is in line with the trend of neutron yield with respect to the incident laser intensity reported in literature (see Fig.~\ref{scaling}(a)). In this case the neutrons are generated either by the thermonuclear reactions in the hot dense plasma produced by the laser interaction, or by the fusion reaction in the target bulk driven by the ions accelerated at the laser front surface through the hole-boring mechanism~\cite{hole_boring}. 

\begin{figure}
\includegraphics[angle=-90,width=0.48\textwidth]{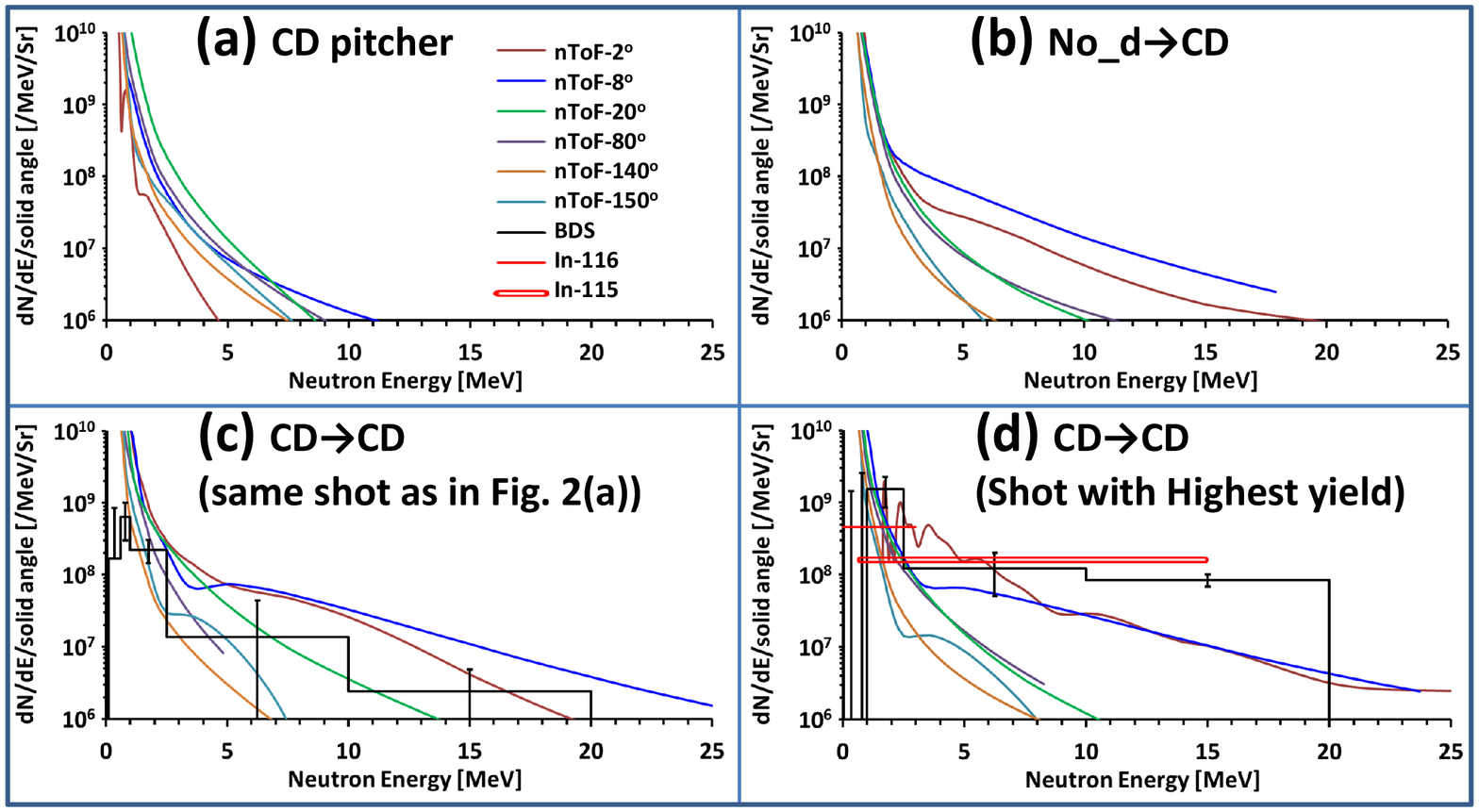}
\caption{\textbf{(a)}-\textbf{(d)} show the comparison between angularly resolved neutron spectra obtained by the six scintillator detectors (shown in Fig.~\ref{Setup}) obtained in different pitcher-catcher combinations as labelled on the top of the graph. The graph also show the data obtained from BDS and activation diagnostics, where available. Where BDS and activation diagnostics were installed along the beam axis in case of (d), as shown in fig.~\ref{Setup}, the BDS in (c) was installed at 145 deg from laser axis.}
\label{spectra}
\end{figure}

The isotropic neutron emission from pitcher-only targets, which is commonly observed in experiments as shown in Fig.~\ref{scaling}(b), is in stark contrast to the forwardly-directed, beamed neutron flux obtained in pitcher-catcher configuration as shown in Fig.~\ref{scan} and Fig.~\ref{spectra}(a). As can be seen in Fig.~\ref{scaling}(b), our data represents the highest degree of beam anisotropy observed experimentally with peak flux of the order of $10^9$ n/sr. 
The order of magnitude increase in neutron flux along the ion beam axis (which is same as the incident laser axis) observed in this case originates from several different nuclear reactions inside the catcher. Possible reactions include the interaction of protons, deuteron and carbon ions produced from the pitcher target by the TNSA mechanism with the carbon and deuterium atoms in the catcher. 

\begin{figure}
\includegraphics[angle=-90,width=0.48\textwidth]{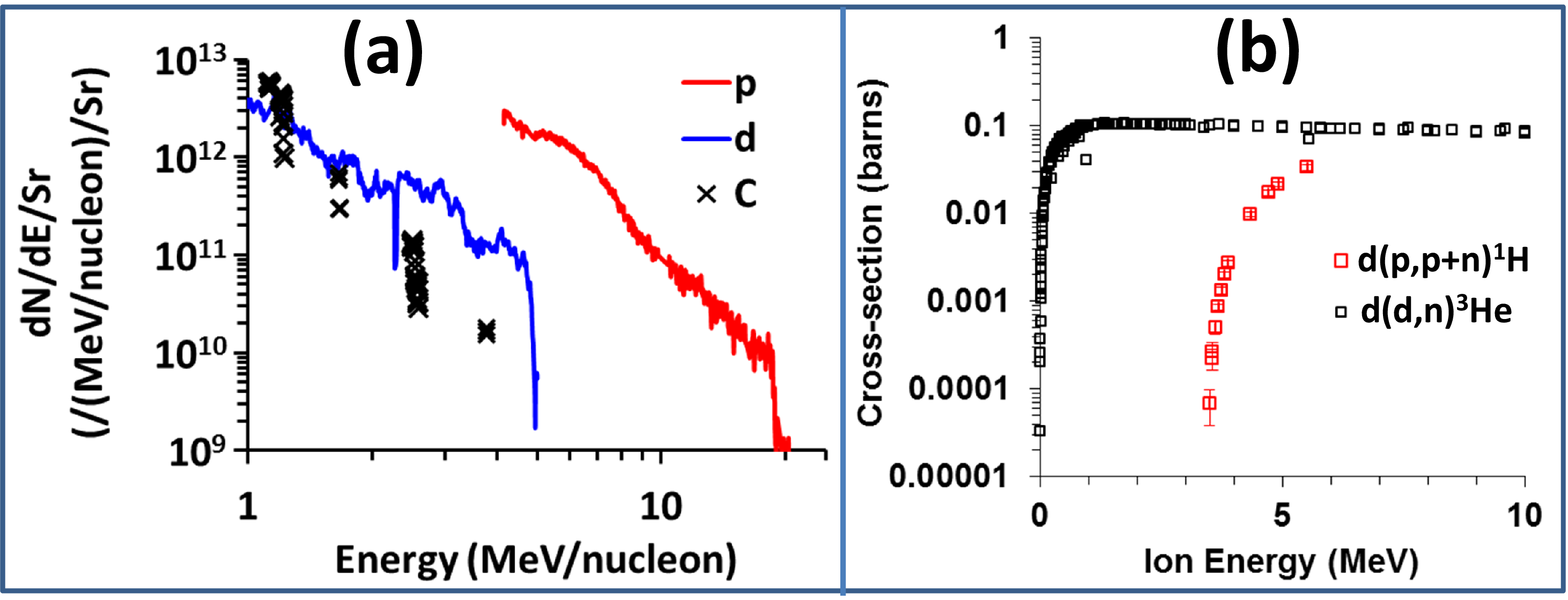}
\caption{\textbf{(a)} Comparison between on axis proton (red), deuteron (blue) and carbon (black) ion spectra obtained from 10 $\um$ thick CD target. \textbf{(b)} Comparison between reaction cross-sections for d(p, n+p)\textsuperscript{1}H (red) and d(d,n)\textsuperscript{3}He (black) for different projectile ion (protons and deuterons respectively)) energies, as obtained from EXFOR~\cite{exfor}.}
\label{ions}
\end{figure}

Different pitcher-catcher combinations 
were used in order to unfold the contributions from different potential reactions towards the observed beamed flux of neutrons. Firstly, employing graphite catcher with CD pitcher target (CD$\rightarrow$C) did not produce (in spite of shot-to-shot fluctuations) a significant increase in the neutron flux or beam anisotropy compared to pitcher-only target. This suggests insignificant contribution being made by the p+C, d+C and C+C reactions in the context of the observed neutron flux in the CD$\rightarrow$CD case. Secondly, in order to discriminate between proton and deuteron induced reactions in the CD targets, several shots were taken by using a CD catcher in front of a deuterium free ion source (No$\_$d$\rightarrow$CD) by using either 10$\um$ gold or CH foils as pitcher targets (Au/CH$\rightarrow$CD), or 10$\um$ CD foil backed by few microns thick aluminium foil as pitcher targets (CD+Al$\rightarrow$CD). 
Similar neutron flux as in the CD$\rightarrow$CD case was observed in this case (No$\_$d$\rightarrow$CD). 
Since the deuteron and carbon ion spectra produced from the pitcher targets are similar in terms of number of particles and beam temperature, as shown in Fig.~\ref{ions}(a), d+C reactions in case of No$\_$d$\rightarrow$CD can also be assumed insignificant.
Therefore, the most promising reaction in this case would be the proton driven neutron generation via breakup of deuterons (d(p, n+p)\textsuperscript{1}H) in the catcher. This is expected due to (1) large number of high energy protons ($\sim$20 MeV) being produced from the pitcher target by the TNSA mechanism, as shown in Fig.~\ref{ions}(a), and (2) high cross-section of d(p, n+p)\textsuperscript{1}H reaction (although there is a limited amount of data available~\cite{exfor}) as shown in Fig.~\ref{ions}(b). Since cleaning techniques for removal of target contaminants (such as laser ablation, resistive heating etc.) could not be implemented in the experiment due to setup constraints, it was also not possible to study the interaction of a proton-free ion beam with CD catchers. However, by comparing the data obtained from No$\_$d$\rightarrow$CD and CD$\rightarrow$CD cases in Fig.~\ref{scan}, one can reasonably assume a similar contributions from d(d, n)\textsuperscript{3}He and d(p, n+p)\textsuperscript{1}H reactions in the CD$\rightarrow$CD case. 

Angularly resolved neutron spectra obtained by the six nToF detectors, as shown in Fig.~\ref{spectra}, show the difference in neutron generation between the cases. While the CD$\rightarrow$C case produced low energy neutrons isotropically, similar to that obtained for pitcher only target shown in Fig.~\ref{spectra}(a), No$\_$d$\rightarrow$CD and CD$\rightarrow$CD produced a very significant anisotropy in both neutron flux and maximum neutron energy, as shown in Fig.~\ref{spectra}(b) and (c) respectively. Fig.~\ref{spectra}(c) shows the neutron spectra obtained along different angles for the case shown in Fig.~\ref{cr39}(a), where the spectrally integrated neutron flux over the neutron energy in the range 2.5-10 MeV obtained from the nToF detectors agrees well within the experimental errors with the off axis BDS and CR39 measurements shown in Fig.~\ref{cr39}(e). The highest neutron flux obtained in the experiment was for the CD$\rightarrow$CD case, where, as shown in the Fig.~\ref{spectra}(d), the nToF spectra agrees with that obtained from the BDS and activation diagnostics. The peak neutron flux along the beam axis, for neutron energy between 2.5-10 MeV, was close to 10$^9$ n/sr. Indeed, the neutron flux can be significantly enhanced by optimising the neutron generation with high yield catcher targets, such as lithium or beryllium, for which the reaction cross-section is an order of magnitude higher than for d(d,n)\textsuperscript{3}He reaction and stays very high for ions with tens of MeV energy.

The anisotropy ratios (which we define as the ratio of on-axis flux to off-axis flux) that have been observed in these experiments can be estimated for a beam-fusion scenario in terms of the projectile energy and the differential cross-section of the nuclear reaction. For instance, in case of d(d,n)$^3$He reaction, using the ion velocity in the centre-of-mass (c-o-m) frame, $v_{d,cm} = (1/2)\sqrt{E_{d}/2m_p}$, one can obtain the emitted neutron velocity in the c-o-m frame from the energy-momentum conservation as $v_{n,cm}=\sqrt{3(E_{d}+Q)/4m_p}$, where $E_{d}$ is the incident deuteron energy in the laboratory (lab) frame, $m_p$ is the mass of proton and $Q$ is the Q-value of the reaction. Therefore, the neutron energy in the lab frame along a given neutron emission angle ($\theta$, with respect to the incident ion beam axis) can be written as
\begin{equation}
E_{n} = E_{d}/8\left[\sqrt{\cos^2\theta+2+6Q/E_{d}}+\cos\theta \right]^2.
\label{E_n_eq}
\end{equation}  
From this expression one can determine the neutron velocity in the lab frame $v_{n}$, and thus the neutron emission angle in the c-o-m frame can be written as $\cos\theta_{cm} = (v_{n}\cos\theta - v_{d})/v_{n,cm}$. The anisotropy is then determined from the differential cross-section in the lab frame, which is related to that in the c-o-m frame via,
\begin{equation}
\frac{d\sigma}{d\Omega} =\frac{(1+\alpha^2+2\alpha\cos\theta_{cm})^{3/2}}{1+\alpha\cos\theta_{cm}}\left[\frac{d\sigma}{d\Omega} \right]_{cm}.
\label{aniso}
\end{equation} 
where, $\alpha=v_{d,cm}/v_{n,cm}$. An anisotropy ratio can thus be obtained at each deuteron energy by taking the ratio of the differential cross-section values that are calculated in this way at 0 and 90 degrees. This requires one to use the tabulated data for the differential cross-section in the c-o-m frame available in the Experimental Nuclear Reaction Database(EXFOR)~\cite{exfor}. 

\begin{figure}
\includegraphics[angle=-90, width=0.48\textwidth]{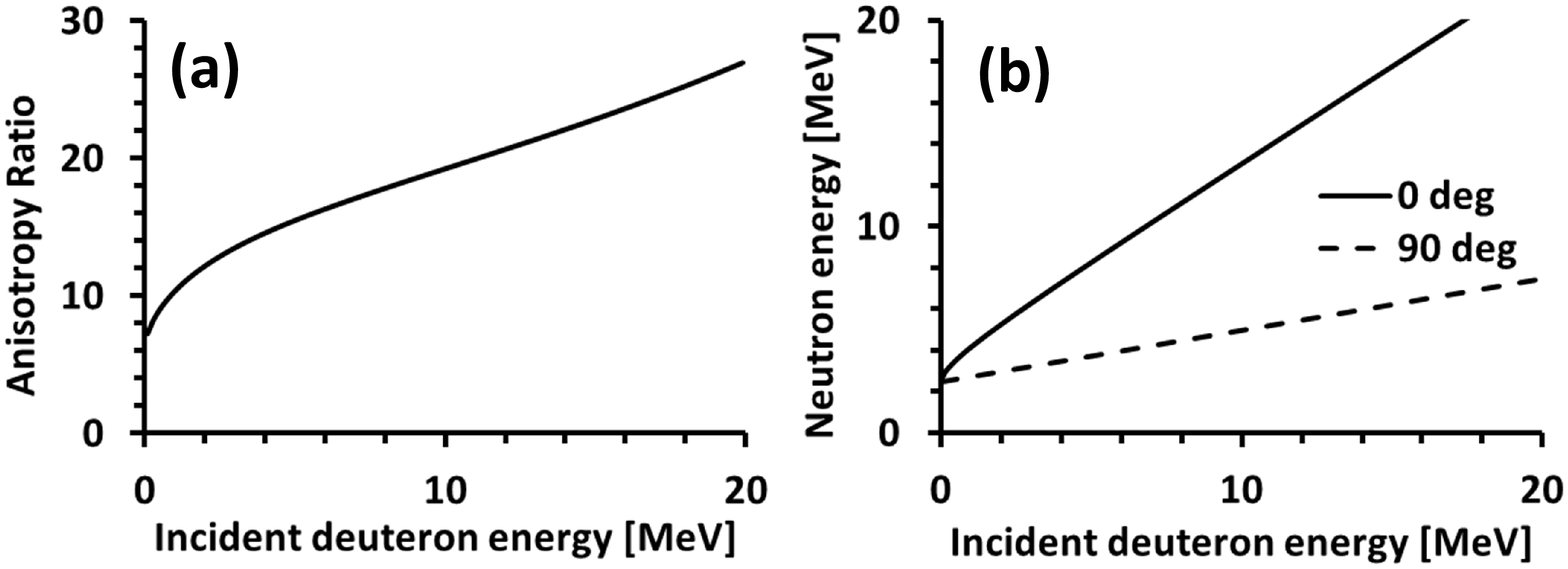}
\caption{\textbf{(a)} Theoretical anisotropy ratio for d(d,n)$^3$He reactions as a function of incident deuteron energy using Eq.~\ref{aniso}. \textbf{(b)} Energy of emitted neutrons in a beam-fusion scenario along 0 and 90 degrees to the incident ion beam axis as a function of incident deuteron energy, calculated using Eq.~\ref{E_n_eq}.}
\label{fig:aniso}
\end{figure}

The results of this calculation are shown in Fig.\ref{fig:aniso}(a), which indicates that the d(d,n)$^3$He reactions could easily produce the levels of anistropy shown in Fig.\ref{scaling} given that the deuteron energy spectrum shown in Fig.\ref{ions} extends above 5~MeV. The anisotropy emission from this calculation assumes that the angular distribution of the incident ion beam is sufficiently narrower than that of the neutrons. Without the differential cross-sections for d(p, n+p)\textsuperscript{1}H reactions in the c-o-m frame, it is hard to produce a similar graph for the d(p, n+p)\textsuperscript{1}H reaction. However, even if one assumes an isotropic emission of neutrons for d(p, n+p)\textsuperscript{1}H reaction in the c-o-m frame, the kinematic effects still lead to an anisotropy ratio of about 10 over a wide range of energies. 

In addition to the observed anisotropy in neutron flux, the energy distribution arising from the reactions, as can be seen from Fig.~\ref{spectra}, is highly anisotropic due to nuclear reaction kinematics. 
According to Eq.~\ref{E_n_eq}, one can approximate the energy of emitted neutrons along the beam forward direction (see Fig.~\ref{fig:aniso}(b)) as $E_n(\theta=0) \approx E_{d}$ while $E_d > Q$ (3.3 MeV and -2.25 MeV for d(d,n)$^3$He and d(p, n+p)\textsuperscript{1}H reactions respectively), which is similar to the measured neutron energy by the nToF detectors as shown in Fig.~\ref{spectra}. 



In conclusion, we have demonstrated a strongly beamed ($\sim70\,^{\circ}$ FWHM), high flux (of the order of $10^9$ n/sr) source of fast neutrons based on beam-nuclear reaction employing high power laser driven protons and deuterons. The neutron flux in the beam, which was amongst the highest reported in the literature, was order of magnitude higher than that present elsewhere, as characterised spatially and spectrally by a suite of neutron diagnostics. Such a directed beam of fast neutrons is highly favorable not only for its direct applications, but also for its transport and moderation. 
With the possibility of producing higher energy ion beams and higher flux, either by TNSA or via emerging ion acceleration mechanisms, and by employing higher yield neutrons converters (such as \textsuperscript{7}Li or \textsuperscript{9}Be), this approach can lead to development of an appealing neutron source for both established and innovative applications. 

The authors acknowledge funding from EPSRC, [EP/J002550/1-Career Acceleration Fellowship held by S.K., EP/L002221/1, EP/K022415/1, EP/J500094/1, EP/J003832/1 and EP/I029206/1], SBF-TR18 and GRK1203 projects. Authors acknowledge support of engineering, target fabrication and experimental science groups of Central Laser Facility of STFC, UK. 



\begin{thebibliography}{99}

\bibitem{Perkins} L.J. Perkins \textit{et al.}, Nucl. Fusion, \textbf{40}, 1 (2000); K. Ehrlich and A. Moslang, Nucl. Instru. Methods B, \textbf{139}, 72 (1998); A. Moslang, Comptes rendus Physique, \textbf{9}, 457 (2008).

\bibitem{NRS} A. Widenmann \textit{et al.}, Phys. Rev. Lett., \textbf{97}, 057202, (2006); D. P. Higginson \textit{et al.}, Phys. Plasmas \textbf{17}, 100701 (2010).

\bibitem{PFNA1} J. Rynes \textit{et al.}, Nucl. Instr. Meth. Phys. Res. A \textbf{422} 895 (1999); B.D. Sowerby and J.R. Tickner, Nucl. Instr. Meth. Phys. Res. A \textbf{580} 799 (2007); H.Y. Lu \emph{et al.}, Phys. Rev. A, 80, 051201 (2009); J.S. Brzosko \textit{et al.}, Nucl. Instrum. Methods B, \textbf{72}, 119 (1992); R. Loveman \textit{et al.}, Nucl. Instrum. Methods B, \textbf{99}, 765 (1995)

\bibitem{Roth_Neutron} M. Roth \textit{et al.}, Phys. Rev. Lett. \textbf{110}, 044802 (2013)

\bibitem{RPA} APL Robinson \textit{et al.}, Plasma Phys. Control. Fusion, \textbf{51} 024004 (2009); S. Kar \textit{et al.}, Phys. Rev. Lett. \textbf{100}, 225004 (2008); S. Kar \textit{et al.}, Phys. Rev. Lett., \textbf{109}, 185006 (2012).

\bibitem{BOA} L. Yin \textit{et al.}, Phys. Plasmas \textbf{14}, 056706 (2007)

\bibitem{TNSA} J. Fuchs \textit{et al.}, Nat. Phys. \textbf{2}, 48 (2006); L. Robson \textit{et al.}, Nature Phys., \textbf{3} , 58 (2006).

\bibitem{Petrov_sim}J. Davis and G.M. Petrov, Plasma Phys. Control. Fusion, 50, 065016 (2008); J. Davis and G.M. Petrov, Plasma Phys. Control. Fusion, 52, 045015 (2011);

\bibitem{isis} http://www.isis.stfc.ac.uk/; http://europeanspallationsource.se/; http://neutrons.ornl.gov/sns.

\bibitem{bnct} A. Wittig \textit{et al.}, Crit. Rev. Oncology/Hematology, \textbf{68}, 66 (2008); R.Terlizzi \textit{et al.},Appl. Rad. Isotopes, \textbf{67}, S292 (2009); 

\bibitem{Williangle_2011} L. Williangle \textit{et al.}, Phys. Plasmas, \textbf{18}, 083106 (2011)

\bibitem{Zulick_2013} C. Zulick \textit{et al.}, Appl. Phys. Letts., \textbf{102}, 124101 (2013)

\bibitem{Deborah_TP} 
D. Gwynne \textit{et al.}, Rev. Sci. Instrum. \textbf{85}, 033304 (2014)

\bibitem{Aaron_TP} 
A. Alejo \textit{et al.}, Rev. Sci. Instrum. \textbf{85}, 093303 (2014)

\bibitem{srim} www.srim.org

\bibitem{Frenje} J. A. Frenje \textit{et al.}, Rev. Sci. Instrum. \textbf{73}, 7 (2002)

\bibitem{Bubbles} http://bubbletech.ca/; A. Green \emph{et al.}, Central Laser Facility Annual report, p. 73, 2012-13.

\bibitem{activation} S. Dorkings \textit{et al.}, Central Laser Facility Annual report, p. 75, 2012-13.

\bibitem{ntof_calibration} S. Mirfyazi \textit{et al.}, arxiv:1506.04689.


\bibitem{Habara_2004} H. Habara \textit{et al.}, Phys. Rev. E, \textbf{70}, 046414 (2004)

\bibitem{Disdier_1999} L. Disdier \textit{et al.}, Phys. Rev. Lett., \textbf{82}, 1454 (1999)

\bibitem{Izumi_2002} N. Izumi \textit{et al.}, Phys. Rev. E, \textbf{65}, 036413 (2002)

\bibitem{Norreys_1998} P. Norreys \textit{et al.}, Plasma Phys. Control. Fusion, \textbf{40}, 175 (1998)

\bibitem{Higginson_2011} D.P. Higginson \textit{et al.}, Phys. Plasmas, \textbf{18}, 100703 (2011)

\bibitem{Yang_JAP_2004} J.M. Yang~\textit{et al.}, J. App. Phys., \textbf{96}, 6912 (2004).

\bibitem{Lancaster_2004} K.L. Lancaster \textit{et al.}, Phys. Plasmas, \textbf{11}, 3404 (2004)

\bibitem{Habara_2004_2} H. Habara \textit{et al.}, Phys. Rev. E, \textbf{69}, 036407 (2004)

\bibitem{hole_boring} APL. Robinson \textit{et al.}, Plasma Phys. Control. Fusion, \textbf{51}, 024004 (2009); S. Kar \textit{et al.}, Phys. Rev. Lett. \textbf{100}, 225004 (2008)

\bibitem{exfor} https://www-nds.iaea.org/exfor/exfor.htm

\bibitem{plasma_mirror} Ch. Ziener  \textit{et al.}, J. Appl. Phys., \textbf{93}, 768 (2003); B. Dromey \textit{et al.}, Rev. Sci. Instrum., \textbf{75}, 645 (2004); 







\end{thebibliography}
\end{document}